\documentstyle[12pt,epsf]{article}
\begin{document}

\catcode`@=11
\long\def\@caption#1[#2]#3{\par\addcontentsline{\csname
  ext@#1\endcsname}{#1}{\protect\numberline{\csname
  the#1\endcsname}{\ignorespaces #2}}\begingroup
    \small
    \@parboxrestore
    \@makecaption{\csname fnum@#1\endcsname}{\ignorespaces #3}\par
  \endgroup}
\catcode`@=12
\newcommand{\newc}{\newcommand}
\newc{\gsim}{\lower.7ex\hbox{$\;\stackrel{\textstyle>}{\sim}\;$}}
\newc{\lsim}{\lower.7ex\hbox{$\;\stackrel{\textstyle<}{\sim}\;$}}
\newc{\gev}{\,{\rm GeV}}
\newc{\mev}{\,{\rm MeV}}
\newc{\ev}{\,{\rm eV}}
\newc{\kev}{\,{\rm keV}}
\newc{\tev}{\,{\rm TeV}}
\def\tr{\mathop{\rm tr}}
\def\Tr{\mathop{\rm Tr}}
\def\Im{\mathop{\rm Im}}
\def\Re{\mathop{\rm Re}}
\def\bR{\mathop{\bf R}}
\def\bC{\mathop{\bf C}}
\def\lie{\mathop{\hbox{\it\$}}} 
\newc{\sw}{s_W}
\newc{\cw}{c_W}
\newc{\swsq}{s^2_W}
\newc{\swsqb}{s^2_W}
\newc{\cwsq}{c^2_W}
\newc{\cwsqb}{c^2_W}
\newc{\Qeff}{Q_{\rm eff}}
\newc{\fpf}{{\bf\bar5}+{\bf5}}
\newc{\tpt}{{\bf\overline{10}}+{\bf10}}
%
%
\def\NPB#1#2#3{Nucl. Phys. {\bf B#1} (19#2) #3}
\def\PLB#1#2#3{Phys. Lett. {\bf B#1} (19#2) #3}
\def\PLBold#1#2#3{Phys. Lett. {\bf#1B} (19#2) #3}
\def\PRD#1#2#3{Phys. Rev. {\bf D#1} (19#2) #3}
\def\PRL#1#2#3{Phys. Rev. Lett. {\bf#1} (19#2) #3}
\def\PRT#1#2#3{Phys. Rep. {\bf#1} (19#2) #3}
\def\ARAA#1#2#3{Ann. Rev. Astron. Astrophys. {\bf#1} (19#2) #3}
\def\ARNP#1#2#3{Ann. Rev. Nucl. Part. Sci. {\bf#1} (19#2) #3}
\def\MPL#1#2#3{Mod. Phys. Lett. {\bf #1} (19#2) #3}
\def\ZPC#1#2#3{Zeit. f\"ur Physik {\bf C#1} (19#2) #3}
\def\APJ#1#2#3{Ap. J. {\bf #1} (19#2) #3}
\def\AP#1#2#3{{Ann. Phys. } {\bf #1} (19#2) #3}
\def\RMP#1#2#3{{Rev. Mod. Phys. } {\bf #1} (19#2) #3}
\def\CMP#1#2#3{{Comm. Math. Phys. } {\bf #1} (19#2) #3}
\relax
%
%
%
\def\beq{\begin{equation}}
\def\eeq{\end{equation}}
\def\bea{\begin{eqnarray}}
\def\eea{\end{eqnarray}}
%
%
%
\def\boxeqn#1{\vcenter{\vbox{\hrule\hbox{\vrule\kern3pt\vbox{\kern3pt
\hbox{${\displaystyle #1}$}\kern3pt}\kern3pt\vrule}\hrule}}}
%
%
\def\mbox#1#2{\vcenter{\hrule \hbox{\vrule height#2in
\kern#1in \vrule} \hrule}}
\def\half{{\textstyle{1\over2}}} 
%
%
%
%
\newc{\ie}{{\it i.e.}}          \newc{\etal}{{\it et al.}}
\newc{\eg}{{\it e.g.}}          \newc{\etc}{{\it etc.}}
\newc{\cf}{{\it c.f.}}
%
%
%
%
\def\CAG{{\cal A/\cal G}} 
\def\CA{{\cal A}} \def\CB{{\cal B}} \def\CC{{\cal C}} \def\CD{{\cal D}}
\def\CE{{\cal E}} \def\CF{{\cal F}} \def\CG{{\cal G}} \def\CH{{\cal H}}
\def\CI{{\cal I}} \def\CJ{{\cal J}} \def\CK{{\cal K}} \def\CL{{\cal L}}
\def\CM{{\cal M}} \def\CN{{\cal N}} \def\CO{{\cal O}} \def\CP{{\cal P}}
\def\CQ{{\cal Q}} \def\CR{{\cal R}} \def\CS{{\cal S}} \def\CT{{\cal T}}
\def\CU{{\cal U}} \def\CV{{\cal V}} \def\CW{{\cal W}} \def\CX{{\cal X}}
\def\CY{{\cal Y}} \def\CZ{{\cal Z}}
%
%
%
%
%
\def\grad#1{\,\nabla\!_{{#1}}\,}
\def\gradgrad#1#2{\,\nabla\!_{{#1}}\nabla\!_{{#2}}\,}
\def\partder#1#2{{\partial #1\over\partial #2}}
\def\secder#1#2#3{{\partial^2 #1\over\partial #2 \partial #3}}
%
%
%
%
%
\def\ltap{\ \raise.3ex\hbox{$<$\kern-.75em\lower1ex\hbox{$\sim$}}\ }
\def\gtap{\ \raise.3ex\hbox{$>$\kern-.75em\lower1ex\hbox{$\sim$}}\ }
\def\gl{\ \raise.5ex\hbox{$>$}\kern-.8em\lower.5ex\hbox{$<$}\ }
\def\roughly#1{\raise.3ex\hbox{$#1$\kern-.75em\lower1ex\hbox{$\sim$}}}
%
%
%
%
\def\slash#1{\rlap{$#1$}/} 
\def\dsl{\,\raise.15ex\hbox{/}\mkern-13.5mu D} 
\def\delsl{\raise.15ex\hbox{/}\kern-.57em\partial}
\def\Ksl{\hbox{/\kern-.6000em\rm K}}
\def\Asl{\hbox{/\kern-.6500em \rm A}}
\def\Dsl{\hbox{/\kern-.6000em\rm D}} 
\def\Qsl{\hbox{/\kern-.6000em\rm Q}}
\def\gradsl{\hbox{/\kern-.6500em$\nabla$}}
%
%
\let\al=\alpha
\let\be=\beta
\let\ga=\gamma
\let\Ga=\Gamma
\let\de=\delta
\let\De=\Delta
\let\ep=\varepsilon
\let\ze=\zeta
\let\ka=\kappa
\let\la=\lambda
\let\La=\Lambda
\let\del=\nabla
\let\si=\sigma
\let\Si=\Sigma
\let\th=\theta
\let\Up=\Upsilon
\let\om=\omega
\let\Om=\Omega
\def\ph{\varphi}
%
%
%
\newdimen\pmboffset
\pmboffset 0.022em
\def\oldpmb#1{\setbox0=\hbox{#1}%
 \copy0\kern-\wd0
 \kern\pmboffset\raise 1.732\pmboffset\copy0\kern-\wd0
 \kern\pmboffset\box0}
\def\pmb#1{\mathchoice{\oldpmb{$\displaystyle#1$}}{\oldpmb{$\textstyle#1$}}
	{\oldpmb{$\scriptstyle#1$}}{\oldpmb{$\scriptscriptstyle#1$}}}
%
%
%
%
%
\def\bar#1{\overline{#1}}
\def\vev#1{\left\langle #1 \right\rangle}
\def\bra#1{\left\langle #1\right|}
\def\ket#1{\left| #1\right\rangle}
\def\abs#1{\left| #1\right|}
\def\vector#1{{\vec{#1}}}
\def\inv{^{\raise.15ex\hbox{${\scriptscriptstyle -}$}\kern-.05em 1}}
\def\pr#1{#1^\prime}  
\def\lbar{{\lower.35ex\hbox{$\mathchar'26$}\mkern-10mu\lambda}} 
\def\e#1{{\rm e}^{^{\textstyle#1}}}
\def\ee#1{\times 10^{#1} }
\def\om#1#2{\omega^{#1}{}_{#2}}
\def\imp{~\Rightarrow}
\def\coker{\mathop{\rm coker}}
\let\p=\partial
\let\<=\langle
\let\>=\rangle
\let\ad=\dagger
\let\txt=\textstyle
\let\h=\hbox
\let\+=\uparrow
\let\-=\downarrow
\def\dot{\!\cdot\!}
\def\vfilll{\vskip 0pt plus 1filll}
%

\begin{titlepage}
\begin{flushright}
{\large
IASSNS-HEP-96/57\\
hep-ph/9606215\\
May 1996\\
}
\end{flushright}
\vskip 1cm
\begin{center}
{\Large\bf The problems of unification--mismatch and low $\alpha_3$:}\\ \vskip .2cm
{\Large\bf A solution with light vector--like matter\footnote{Research
supported in part by DOE grant DE-FG02-90ER40542 and in part by NSF
grant No. PHY-9119745.
Email: {\tt babu@sns.ias.edu, pati@umdhep.umd.edu}}}
\vskip 1cm
{\large
K.S.~Babu$^*$
and Jogesh C. Pati$^\dagger$\\}
\vskip 0.5cm
{\large\sl $^*$School of Natural Sciences\\
Institute for Advanced Study\\
Princeton, NJ~08540\\} \vskip .3cm
{\large\sl $^\dagger$Department of Physics\\
University of Maryland\\
College Park, MD~20742\\}
\end{center}
\vskip .5cm
\begin{abstract}

The commonly accepted notion of a weak unified coupling $\alpha_X \approx
0.04$, based on the assumption of the MSSM--spectrum, is questioned.  
It is suggested
that the four--dimensional unified string coupling should very likely have
an intermediate value $(\sim 0.2-0.3$, say) so that it may be large
enough to stabilize the dilaton but not so large as to disturb the
coupling--unification relations.  Bearing this in mind, as well as the
smallness of the MSSM unification scale $M_X$ compared to the string
scale, the consequences of a previously suggested extension of the
MSSM spectrum are explored.  The extension contains two vector--like
families of quarks and leptons with relatively light masses of order 1 TeV, having
the quantum numbers of ${\bf 16}+{\bf \overline{16}}$ of $SO(10)$.
It is observed that such an extension provides certain unique
advantages.  These include: (a) removing the stated mismatch between
MSSM and string unifications with
regard to $\alpha_X$ and to some extent $M_X$ as well, (b) achieving
coupling unification with a relatively low value of $\alpha_3(m_Z)$,
in accord with its world average value, and (c) following earlier works,
providing a simple explanation of the observed inter--family
mass--hierarchy.  The extension provides scope for exciting new
discoveries, beyond those of SUSY and Higgs particles, at future
colliders, including the LHC and the NLC.

\end{abstract}
\end{titlepage}
\setcounter{footnote}{0}
\setcounter{page}{1}
\setcounter{section}{0}
\setcounter{subsection}{0}
\setcounter{subsubsection}{0}

\baselineskip= .30in
\section{Introduction} \label{sec:intro}

Achieving a complete unity of the fundamental forces together with an 
understanding of the origin of the three families and their
hierarchical masses is among the major challenges still confronting
particle physics.  Conventional grand unification  falls
short in this regard in that owing to the arbitrariness in the Higgs
sector, it does not unify the Higgs exchange force, not to
mention gravity.  Superstring theory  is the only theory
we know that seems capable of removing these shortcomings.  It thus
seems imperative that the low energy data extrapolated to high
energies be compatible with string unification.  

It is, however, known \cite{keith} that while the three gauge couplings,
extrapolated in the context of the minimal supersymmetric standard
model (MSSM) meet, at least aproximately [2-5], provided $\alpha_3(m_Z)$ is
not too low (see later), their scale of meeting,
$M_X \approx 2 \times 10^{16}~GeV$, is nearly 20 times smaller than
the expected (one--loop level) string--unification scale \cite{ginsparg} of
$M_{\rm st} \simeq g_{\rm st} \times (5.2 \times 10^{17}~GeV) \simeq
3.6 \times 10^{17}~GeV$.  

It seems to us that there is still 
a second mismatch concerning the value of the
unified gauge coupling $\alpha_X$ at $M_X$.  Subject to the assumption
of the MSSM spectrum, extrapolation
of the low energy data yields a rather low value of $\alpha_X \sim 0.04$ [2-5],  
for which perturbative physics should work well near $M_X$.  
On the other hand, it is known \cite{dineseiberg}
that non--perturbative physics ought to be important for a
string theory near the string  scale, in order
that it may help choose the true vacuum and fix the moduli and the
dilaton VEVs.  The need to stabilize the dilaton in particular would suggest 
that the value of the unified
coupling at $M_{\rm st}$ in four dimensions should be considerably
larger than $0.04$ \cite{pert}.  At the same
time, $\alpha_{\rm st}$ should not be too large, because, if
$\alpha_{\rm st} \gg 1$, the corresponding theory should be equivalent
by string duality \cite{dual} to a certain
weakly coupled theory that would still suffer from the dilaton runaway
problem \cite{dine}.  Furthermore, $\alpha_{\rm st}$ at $M_{\rm st}$ should not
probably be as large as even unity, or else, the one--loop string
unification relations for the gauge couplings \cite{ginsparg} would cease to
hold near $M_{\rm st}$ (e.g. in this case, the string threshold
corrections are expected to be too large) and the observed
(approximate) meeting of the three couplings would have to be viewed
as an accident.  In balance, therefore, the preceding discussions 
suggest that an
{\it intermediate value} of the string coupling $\alpha_{\rm st} \sim
0.2-0.3$ at $M_{\rm st}$ in four dimensions, which might be large
enough to stabilize the dilaton, but not so large as to disturb
significantly the coupling unification relations, is perhaps the more desired
value.  It is thus a
challenge to find a suitable variant or alternative to MSSM which
removes the mismatch not only with regard to the meeting point $M_X$, but
also with regard to the value of $\alpha_X$.

A third relevant issue is that the world average value of
$\alpha_3(m_Z) = 0.117 \pm 0.005$ \cite{pdg} seems to
be low compared to
its value  that is needed for MSSM unification.  Barring possible
corrections from GUT threshold and Planck scale effects, the
latter is higher than about $0.127$, if $m_{\tilde{q}} < 1~TeV$ and
$m_{1/2} < 500~GeV$ [2-5].  

It is conceivable that the resolution of {\it all three issues} raised
above-i.e. (a) understanding fermion mass--hierarchy, (b) removing incompatibility
between MSSM and string unification,  and (c) accommodating low
$\alpha_3(m_Z)$- have a {\it common denominator}.
The purpose of this note is to explore just this possibility, the
common denominator in question being a previously suggested extension
of the MSSM spectrum \cite{patip,babu,vector} 
that contains two vector--like families and
their SUSY partners, having the quantum numbers of ${\bf 16} +
{\bf \overline{16}}$ of $SO(10)$, all with masses of order 1 TeV.

It has been noted for some time that the existence of two such 
families enables one to obtain a simple understanding of the observed
inter-family mass-hierarchy of the three chiral families \cite{babu}.
The argument will be presented briefly in Sec. 3.  
On the experimental front it is interesting to note that although the
precision measurements of $N_\nu$ and of the oblique electroweak
parameters ($S,T$ and $U$)
disfavor a fourth chiral family, they are rather insensitive to
vector-like families \cite{zhang,vector}.  

The existence of two vector--like families together with three chiral
families and the associated form of
the $5 \times 5$ fermion mass matrix was in fact derived 
in the context of a SUSY preon model \cite{patip,hanns}.   Such a spectrum could 
well emerge, however, even if quarks, leptons and Higgs bosons are
elementary, e.g. from a superstring theory.  
In view of its prospects for providing exciting discoveries at the LHC and NLC,
we propose to explore here whether such a spectrum 
might have some additional advantages, {\it in the
context of an elementary quark--lepton--Higgs theory}, in
bridging the gap between MSSM and string unifications
mentioned above, and simultaneously accommodating low $\alpha_3$.  
Before proceeding, we note a few alternative suggestions which have
been proposed to address some of these issues.  

First, a very intriguing suggestion in this regard has recently been
put forth by Witten \cite{ed}.  Using the equivalence of the
strongly coupled heterotic $SO(32)$ and the $E_8 \times E_8$ superstring theories in 
$D=10$, respectively to the weakly coupled $D=10$ Type I and an $M$--theory, 
he observed that the
4-dimensional gauge coupling and $M_{\rm st}$ can both be
small, as suggested by MSSM extrapolation of the low energy data, without
making the Newton's constant unacceptably large.  While this observation opens
up a new perspective on string unification, its precise use to make
$\alpha_{\rm st} \approx 0.04$ at $M_{\rm st}$ would seem to run into
the dilaton runaway problem as in fact noted in Ref. \cite{ed}.  Furthermore,
lowering $M_{\rm st}$ to $2 \times 10^{16}~GeV$ would mean that the 
heavy string
states, very likely including color triplets, would have masses $\sim 2 \times
10^{16}~GeV$.  Generically, this might lead to the problem of
rapid proton decay through dimension 5 operators.  The case of larger
$\alpha_X$ and $M_X$ proposed here (see later) would seem to fare better in
overcoming these potential difficulties.  

A second way in which the mismatch between $M_X$ and $M_{\rm st}$
could be resolved is if superstrings yield an intact grand unification
symmetry like $SU(5)$ or $SO(10)$ with the right spectrum -- i.e.,
three chiral families and a suitable Higgs system including an adjoint
Higgs at $M_{\rm st}$, and if this symmetry
would break spontaneously at $M_X \approx (1/20~ {\rm to}~ 1/50)
M_{\rm st}$ to the standard model symmetry.  However, as yet, there is
no realistic string--derived GUT model \cite{lykken}.
Furthermore, for such solutions, there is the 
likely problem of doublet-triplet splitting and rapid
proton decay. 

A third alternative is based on string--derived standard
model--like gauge groups and attributes the mismatch between $M_X$ and
$M_{\rm st}$ to the existence of new matter with intermediate
scale masses ($\sim 10^9-10^{13}$ GeV), which may emerge from
strings \cite{far}.  Such a resolution is
in principle possible, but it would rely on the delicate balance
between the shifts in the three couplings and on the existence of very
heavy new matter which in practice cannot be directly tested by experiments.
Also, within such alternatives, as well as those based on
non--standard hypercharge normalization \cite{john} and/or large
string--scale threshold effects \cite{nilles}, $\alpha_X$ typically remains
small ($\sim 0.04$), which is not compatible with the
need for a larger $\alpha_X$, as suggested here.  

\section{The Extended Supersymmetric Standard Model (ESSM)}

Bearing in mind the discussions above, we study the running of the
coupling constants within the variant spectrum of quarks and leptons
proposed some time ago \cite{patip,babu,vector} that assumes the
standard model gauge symmetry but
extends the 
MSSM spectrum by adding to it two light
vector-like families $Q_{L,R} = (U,D, N, E)_{L,R}$ and
$Q'_{L,R}=(U',D',N',E')_{L,R}$, two Higgs singlets ($H_{\rm S}$ and $H_\lambda$)
and their SUSY partners, all at about 1 TeV.  We will refer to this
variant as
the Extended Supersymmetric Standard Model (ESSM).  
The combined sets $(Q_L|{\overline{Q_R'}})$ and 
$({\overline{Q_R} }|Q_L'$) transform as ${\bf 16}$ and  ${\bf \overline{16}}$ of
$SO(10)$ respectively.  It is interesting to note that the allowed
extensions of MSSM in the low energy region are rather limited.
Barring addition of singlets, 
ESSM is in fact
{\it the only extension}
of the MSSM, containing complete families of
quarks and leptons, that is permitted by measurements of the oblique
electroweak parameters and $N_\nu$ on the one hand, and
renormalization group analysis on the other hand.  The former
restricts one to add only vector--like (rather than chiral) families 
\cite{zhang},
i.e. only $pairs$ of ${\bf 16} + {\bf \overline{16}}$ of $SO(10)$,
whereas the latter states that no more than one such pair can be added, or else
the gauge couplings would grow too rapidly and would
become nonperturbative far below the unification scale \cite{five}.  
While in this note, we do not address the
derivation of such a spectrum in string theories, it is worth noting
that the emergence of pairs of ${\bf 27} + {\bf \overline{27}}$ of $E_6$ or
${\bf 16} + {\bf \overline{16}}$ of $SO(10)$ in addition to chiral
multiplets is rather generic in string theories \cite{light}.  

Now if the three couplings meet (at least
approximately) for MSSM having 3 famiilies, i.e. three ${\bf 16}$'s, 
at a position $M_X$, they are
guaranteed to meet at the same position {\it in the one-loop approximation} for
ESSM, having an extra pair of ${\bf 16}+{\bf \overline{16}}$.  
But the extra pair having masses $\sim 1~TeV$ 
will inevitably raise the value
of $\alpha_X$ at the meeting point, as desired.  However, they will
not raise $M_X$, in one loop.  But once
$\alpha_X$ is raised to $0.2$ to $0.3$ (see discussions later),
two-loop effects are expected to be important especially near $M_X$.
Our main task thus is to examine whether these two-loop effects for
the ESSM spectrum, including contributions from gauge as well as Yukawa
interactions, would still retain the meeting of the three couplings
while raising $\alpha_X$ as well as $M_X$.

It is worth noting that there have been past attempts \cite{other} to study the
question of the meeting of the coupling constants by adding new
families (chiral or vector) to the MSSM spectrum.  
Our approach and results will
differ, however, from those of the past attempts because (i) We use a
specific (yet most economical) 
pattern of the Yukawa coupling matrix (see below) which is tied to our
desire to understand the inter-family mass hierarchy [12,13].  (ii) We
include the contributions of these Yukawa couplings on the running
of the gauge couplings in two--loop, which turns out to be quite
important, but which have been neglected in past attempts.  
(iii) We use smoothed out threshold effects near the TeV scale [3-5]. 
(iv) And finally, owing to the beneficial effects of the Yukawa couplings 
(see later), we stay
within semi--perturbative limits with $\alpha_X \sim (0.2-0.3)$, 
in contrast to $\alpha_X \sim {\cal O}(1)$ in Ref. \cite{other}
so that our results using the two loop
$\beta$-functions are expected to be more reliable.

\section{The Yukawa coupling matrix in ESSM}

Following Ref. \cite{babu,patip}, it is known that the inter--family mass--hierarchy
is reproduced simply if the three chiral familiies $q_{L,R}^i$, $i=1-3$
derive their masses primarily through their off--diagonal mixings with
the two vector--like families $Q_{L,R}$ and $Q'_{L,R}$.  Short of
deriving such a mass--matrix from a string theory, we will assume
suitable discrete symmetries (see later) which ensure this feature.  
To a 
good approximation the corresponding $5 \times 5$ Yukawa coupling
matrix, near the presumed unification scale, 
is thus assumed to have the simple form:
\begin{eqnarray}
h_{f,c}^{(o)} =
\bordermatrix
{& q_{L}^{i} & Q_{L} & Q_{L}^{\prime}\cr
\overline{q_{R}^{i}} & O & X_fH_{f} & Y_cH_{\rm S}\cr
\overline{Q_{R}} & Y^{\prime\dagger}_cH_{\rm S} & z_cH_{\lambda} & 0\cr
\overline{Q_{R}^{\prime}} & X^{\prime\dagger}_fH_{f} & 0 &
z'_fH_{\lambda}\cr}~. 
\end{eqnarray}

\noindent Here the symbol $q,Q,$ and $Q'$ stand for quarks as well as leptons,
and $i=1,2,3$.  
The subscript $f$ denotes $u,d,l$ or $\nu$, while $c = q$ or $l$ denotes quark
or lepton color.  $H_f$ with $f=u,d$ denotes the familiar two
Higgs doublets, while $H_{\rm S}$ and $H_\lambda$ are Higgs singlet fields.
If the Yukawa couplings satisty left--right, up--down as well as
quark--lepton symmetries at the string scale, we would have $X_f=X_f',
Y_c=Y_c'$ and $z=z'$, and these couplings would be 
independent of flavor and color indices $f$ and
$c$ at that scale.  The zeros appearing in
Eq. (1) are expected to be corrected by terms of order 1 MeV 
through VEVs inserted into
higher dimensional operators.  

The Higgs fields $H_\lambda$, $H_{\rm S}$ and $H_f$ are assumed to acquire VEVs so
that $\left \langle H_\lambda \right \rangle \sim 1~TeV$, $\left \langle
H_{\rm S} \right \rangle \sim \left \langle H_u \right \rangle \sim 250~GeV$
and $\left \langle H_d\right \rangle  \ll \left \langle H_u \right
\rangle$.  To see the reason for family mass hierarchy, though not essential
assume for simplicity $X_f = X'_f$ and $Y_c = Y'_c$ for a moment 
and denote $X^T_f = (x_1,x_2,x_3)_f$ and $Y^T_c = (y_1,y_2,y_3)_c$.  
Regardless of the values of these Yukawa couplings,
one can always rotate the basis vectors so that  $Y^T_c$ is
transformed to the form ${\hat Y}^T_c = (0,0,1)y_c$, ${ X}_f^T$
simultaneously to the form ${\hat X}^T_f = (0,p,1)x_f$, and similarly
$X_f'$ and $Y_c'$.  {\it It is thus
apparent why one family is massless (barring corrections of order 1
MeV), despite lack of any hierarchy in the Yukawa couplings
$(x_1,x_2,x_3)_f$ and $(y_1,y_2,y_3)_c$}; this one
is naturally identified with the electron family.  At the unification
scale one obtains $m_{t,b,\tau}^{(0)} \approx (2x_fy_c)(\left \langle
H_{\rm S} \right \rangle \left \langle H_f \right \rangle/(z\left
\langle H_\lambda \right \rangle))$ and $m_{c,s,\mu}^{(0)} \approx
m_{t,b,\tau}^{(0)} (p^2/4)$.  
A value of
$p \approx (1/4~{\rm to}~1/5)$, which is in the realm of naturalness,
thus provides a big hierarchy of about $(1/64~{\rm to}~1/100)$ between
the masses of the $(c,s,\mu)$ and $(t,b,\tau)$ at the string scale.  
Thus the presence of two vector--like families helps to
provide a simple explanation of the inter--family mass--hierarchy:
$m_{u,d,e} \ll m_{c,s,\mu} \ll m_{t,b,\tau}$ \cite{babu}.  

\section{Renormalization Group Analysis for ESSM}

We have performed a full two--loop analysis of the relevant
renormalization group equations of the gauge couplings including the
contributions of the Yukawa couplings as given in Eq. (1).  To
two--loop order, the RGE for the gauge coupling evolution are given by
\begin{eqnarray}
{d\alpha_i \over dt} = {b_i \over 2\pi} \alpha_i^2 + \sum_{j=1}^3
{b_{ij} \over 8 \pi^2} \alpha_i^2\alpha_j - {\alpha_i^2 \over 2 \pi}
\left({1 \over 16 \pi^2} \right) b_i^{\rm Yuk}
\end{eqnarray}
where the coefficients $b_i$ and $b_{ij}$ are:
\begin{eqnarray}
b_i = \left (\matrix{2 ng + {3 \over 5} n_H \cr -6+2n_g+n_H \cr
-9+2n_g}\right);~
b_{ij} = \left(\matrix{{38 \over 15} n_g + {9\over 25} n_H & {6 \over
5} n_g + {9 \over 5} n_H & {88 \over 15} n_g \cr {2 \over 5} n_g + {3
\over 5} n_H & -24+14 n_g+7n_H & 8n_g \cr {11 \over 15}n_g & 3n_g &
-54+{68 \over 3} n_g}\right)~.
\end{eqnarray}
Here $n_g$ is the total number of generations plus anti--generations
and $n_H$ is the number of $pairs$ of Higgs doublets.  
For the case of ESSM, $n_g=5,n_H=1$, corresponding to 3 chiral and two
vector--like families, and one pair of Higgs doublets $H_u$ and $H_d$.  
The coefficients $b_i^{\rm Yuk}$ appearing in Eq. (2) are given by
\begin{eqnarray}
b_1^{\rm Yuk} &=& {26 \over 5} (x_u'^2+x_u^2)+{14 \over 5}
(x_d'^2+x_d^2)+{18 \over 5} (x_l'^2+x_l^2) + {2 \over 5}(y_q'^2+z_q^2)
\nonumber \\
&~&+ {6 \over 5}(y_l'^2+z_l^2+k_1^2) + {16 \over 5} (z_u'^2+y_u^2)+{4
\over 5}(z_d'^2+y_d^2)+{12 \over 5}(z_l'^2+y_l^2) \nonumber \\
b_2^{\rm Yuk} &=&
6(x_u'^2+x_u^2)+6(x_d'^2+x_d^2)+2(x_l'^2+x_l^2)+6(y_q'^2+z_q^2)+2(y_l'^2
+z_l^2+k_1^2) \nonumber \\
b_3^{\rm Yuk} &=&
4(x_u'^2+x_u^2)+4(x_d'^2+x_d^2)+4(y_q'^2+z_q^2)+2(z_u'^2+y_u^2)+2(z_d'^2+y_d^2)~.
\end{eqnarray}
In addition to the Yukawa couplings given in Eq. (1), we have assumed
the following terms in the superpotential:
\begin{equation}
W \sim k_1 H_uH_dH_\lambda + {k_2 \over 6} H_\lambda^3 + {k_3 \over 6}
H_{\rm S}^3, 
\end{equation}
which gives masses to all the Higgses and Higgsinos and which consists of the
most general set of superpotential terms consistent with a $Z_3 \times
Z_3$ symmetry.  Under this symmetry, which ensures the Yukawa coupling
matrix, Eq. (1), the three chiral families ${\bf
16}_i$ transform as $(\omega,1)$, the vector families transform as
${\bf 16} \sim (1,\omega)$, ${\bf \overline{16}} \sim (\omega,1)$ and
Higgs doublets as $10_H \sim (\omega^2,\omega^2)$ and the Higgs
singlets as $H_\lambda \sim (\omega^2,\omega^2)$, $H_c \sim
(\omega,1)$, where $\omega^3=1$.  

We study the evolution of the gauge couplings using two--loop RGE
(i.e., Eq. (2)) from $m_Z$ upwards by dividing the momentum-range to
two regions:  {\it \underline{Region I}: $(m_Z \le \mu \le \mu_0 \sim 10 M)$:}
Here $M$ denotes the mass of the heaviest particle ($\approx 1-2~TeV$)
in the ESSM spectrum and $\mu_0$ denotes the momentum scale upto which
inclusion of threshold effects is important [3-5].  {\it \underline{Region II}: 
($\mu_0
\le \mu \le 10^{18}~GeV$):} In this region, we treat all particles as
massless.  Taking the couplings at $\mu_0$ as boundary values we use
Eqs. (2)-(4) to extrapolate them upwards.  

For region I, since the masses are spread from $m_Z$ to $M \approx
1.5-2~TeV$, we integrate Eq. (2) piecewise from one threshold ($m_1$)
to the next ($m_2$) by first using the $\theta$--function
approximation for each threshold and using appropriate two--loop
$\beta$--function coefficients ($\tilde{b}_i$ and $\tilde{b}_{ij}$)
for each subregion, which are not exhibited here.  These include contributions
from all particles with masses $\le m_1$ to the evolution of the
couplings in the range $m_1 \le \mu \le m_2$.  Thus ignoring the
contributions from the Yukawa couplings for a moment,
and replacing $b_i$ and $b_{ij}$ in Eq. (2) by $\tilde{b}_i$ and
$\tilde{b}_{ij}$ for the sub--region $m_1 \rightarrow m_2$, Eq. (2)
can be integrated analytically to yield:
\begin{equation}
\tilde{\alpha}_i^{-1}(\mu) = \tilde{\alpha}_i^{-1}(m_1) - {\tilde{b}_i 
\over 2 \pi}{\rm
ln}({\mu \over m_1}) - {1 \over 4 \pi} \sum_j {\tilde{b}_{ij} \over
\tilde{b}_j} {\rm ln}\left[{\tilde{\alpha}_j(\mu) \over
\tilde{\alpha}_j(m_1)}\right]~.
\end{equation}
The contribution of each individual particle denoted by $\hat{b}_i$ to
the regional one--loop coefficients $\tilde{b}_i$ is listed in Table
1.  The corresponding $\tilde{b}_{ij}$ are not exhibited here.
Following this procedure in successive steps (i.e., $m_Z \rightarrow
m_1 \rightarrow m_2 \rightarrow m_3 ....M \rightarrow \mu_0$) we
obtain $\tilde{\alpha}_i^{-1}(\mu_0)$.  Since the leading
log contributions have already been included in Eq. (6), we finally
add to $\tilde{\alpha}_i^{-1}(\mu_0)$ obtained as above, the sum of
the non-logarithmic
threshold corrections for each new 
particle -- ie. $\tilde{\Delta}_i(\mu_0) \equiv
\sum (\Delta_i - {\rm leading~ log~term})$ -- as well as contributions
from the top and the Yukawa couplings of
vector--like quarks to obtain

\begin{equation}
\alpha_i^{-1}(\mu_0) = \tilde{\alpha_i}^{-1}(\mu_0) +
\tilde{\Delta}_i(\mu_0) + \Delta_i^{\rm top} + \Delta_i^{\rm Yuk}
\end{equation}
where $\Delta_i^{\rm top} = (0.138, 0.158,0.090)$ for $m_t = 180~GeV$
[2,3] while $\Delta_i^{\rm Yuk} = (0.026, 0.032, 0.023)$.  To
evaluate $\Delta_i$ and thus $\tilde{\Delta}_i$ we use exact one-loop
threshold functions given by [3,25]
\begin{equation}
\Delta_i^{F,S}(m,\mu_0) = {\hat{b}_i \over 2 \pi} \left[K_{F,S}(m_Z/m)
- K_{F,S}(\mu_0/m)\right] ~.
\end{equation}
\begin{eqnarray}
K_F (q/m) &=& {w^2 \over 2}\left[1-{(w^2-3) \over 2 w}{\rm ln}\left({w+1
\over w-1}\right)\right] \nonumber \\
K_S(q/m) &=& 1-w^2+{1 \over 2} w^3 {\rm ln}\left({w+1 \over w-1}\right)
\end{eqnarray}
Here $(F,S)$ denote (fermion,scalar) 
and $w(q/m) \equiv \sqrt{1+4m^2/q^2}$.

\begin{table}
\centering
\begin{tabular}{|c|ccc|}
\hline
Particles & $\hat{b}_1$ & $\hat{b}_2$ & $\hat{b}_3$ \\
\hline
$\tilde{g}$ & 0 & 0 & 2 \\
$\tilde{W}$ & 0 & ${4 \over 3}$ & 0 \\
$\tilde{B}$ & 0 & 0 & 0 \\
$Q$ & ${1 \over 15}$ & 1 & ${ 2 \over 3}$ \\
$U'$ & ${8 \over 5}$ & 0 & ${1 \over 3}$ \\
$D'$ & ${2 \over 15}$ & 0 & ${1 \over 3}$ \\
$L$ & ${1 \over 5}$ & ${1 \over 3}$ & 0 \\
$E'$ & ${2 \over 5}$ & 0 & 0 \\
$\tilde{l}_L, \tilde{L}$ & ${1 \over 10}$ & ${1 \over 6}$ & 0 \\
$\tilde{q}_L,\tilde{Q}$ & ${1 \over 30}$ & ${1 \over 2}$ & ${1
\over 3}$ \\
$\tilde{u}_R, \tilde{U'}$ & ${4 \over 15}$ & 0 & ${1 \over 6}$
\\
$\tilde{d}_R, \tilde{D'}$ & ${1 \over 15}$ & 0 & ${1 \over 6}$
\\
$\tilde{e}_R,\tilde{E'}$ & ${1 \over 5}$ & 0 & 0 \\
\hline
\end{tabular}
\caption{Threshold function coefficients appearing in Eq. (8) for
various particles in ESSM. $(Q,U',D',L,E')$ are the vector family 
fermions and a tilde denotes SUSY particle.}  
\end{table}

The values of $\Delta_i$'s would depend somewhat, as in MSSM, on the
assumed masses of the new particles.  Considerations based on 
(a) QCD renormalization
effects which enhance the masses of 
($Q,\tilde{Q},\tilde{q},\tilde{g}$) relative to
($L,\tilde{L},\tilde{l}, \tilde{W}$), (b) the need to avoid
unnatural fine--tuning, (so that $m_{\tilde{q}} \le 1~TeV,
|m_Q-m_{\tilde{Q}}| \le 300~GeV$) and (c) 
simplicity of analysis, we 
assume the pattern: $m_Q \approx 1-2~TeV \ge m_{\tilde{Q}} \ge m_L
\sim m_{\tilde {L}} \sim 
m_{\tilde{q}}  \sim m_H \approx m_{\tilde{H}} \ge m_{\tilde{l}}
\ge m_{\tilde{g}} \ge m_{\tilde{W}} \approx 80-200~GeV$.  
The QCD renormalization effects are taken from our
preliminary analysis as a guide, which will be presented eleswhere.
Owing to the added
importance of the two--loop effects in ESSM, even if gaugino masses were
universal at $M_X$, we obtain (ignoring Yukawa effects for this
purpose) $m_{\tilde{g}}/m_{\tilde{W}} \approx 2$.  This is in
contrast to the one--loop value of $m_{\tilde{g}}/m_{\tilde{W}}
\approx \alpha_3/\alpha_2 \approx 3.5$, for MSSM.  
Using this as a rough guide and also allowing for the possible lack of
universality at $M_X$, we will vary $m_{\tilde{g}}/m_{\tilde{W}}$ in
the range of 1.5 to about 3 for ESSM.

To study the evolution of the $\alpha_i$'s in region II ($\mu_0 \le
\mu \le 10^{18}~GeV$), we will assume here that all the relevant
Yukawa couplings involving the third family are large at $M_X$ --i.e.,
$x_i \sim x_i' \sim y_i \sim y_i' \sim z \sim z' \approx 1-2$, so that
they approach their fixed point values near the electroweak scale
\cite{virtue}.  We have derived the full set of one--loop RGE for the
evolution of the Yukawa couplings of the ESSM.  For brevity, these
equations are not presented here \cite{jongbae}.  Solving 
these coupled RGE Eqs.
(2)-(4), and using typical values of $M_X \approx 10^{17}~GeV$ and
$\alpha_X \sim 0.25$ (see later), we find that the Yukawa couplings
acquire their near-fixed point values at 1 TeV, given by:

\begin{eqnarray}
x_u' &=& 0.896, y_q' = 0.746, x_u =0.896, z_q = 0.740, z_u' = 0.554, y_u
= 0.559, \nonumber \\
x_d' &=& 0.871, x_d = 0.872, z_d'=0.533, y_d=0.538, x_l'=0.368,
y_l'=0.251
\nonumber \\
x_l &=& 0.396, z_l=0.273, z_l'=0.185, y_l=0.184, x_\nu'=0.332,
z_\nu'=0.152
\nonumber \\
k_1 &=& 0.010, k_2 = 0.214, k_3=0.217
\end{eqnarray}
These will be taken as their input values at $1~TeV$ \cite{virtue}.  

An interesting comment is in order regarding the value of $m_b/m_\tau$.  
Naively, without the assistance of the Yukawa
couplings, owing to the large ratio $\alpha_3(M_X)/\alpha_3(m_Z)$, $m_b$ would
be much too big compared to experiments at the low scale, if it
were equal to $m_\tau$ at $M_X$.  However, with the effects of the
Yukawa couplings included, we obtain $m_b/m_\tau \simeq
2.53$ at 1 TeV, which is compatible with observation.  

To determine the gauge couplings at $m_Z$ we follow the mass dependent
subtraction procedure (MDSP) \cite{mar}, which is
suited to include the
non--logarithmic threshold effects.  We denote the initial values of
the couplings at $m_Z$ in the MDSP scheme by $\hat{\alpha}_i(m_Z)$ 
\cite{confuse}.  
Following Ref. \cite{bagger,mar}, 
we choose $G_F = 1.6639 \times 10^{-5} GeV^{-2},
m_Z = 91.187~GeV$ and $\alpha^{-1}(0) = 137.036$ as input values
(rather than $\alpha_{em}(m_Z)$ and sin$^2\theta_W(m_Z)$ of the
$\overline{MS}$ scheme), together with a value for $m_t \approx 180~GeV$ and a
chosen ESSM--spectrum to detemrine $\hat{\alpha}_1$ and
$\hat{\alpha}_2$ at $m_Z$.  We next choose a varying input value for
$\hat{\alpha}_3(m_Z) \approx 0.12-0.127$ in the MDSP scheme
\cite{confuse} and
extrapolate the three gauge couplings upward, for a given spectrum, to
test unification.  Following preceding discussions, we
consider a few cases for the spectrum as noted below.  

\underline{{\it Case 1:}}  $m_{\tilde{W}} = 75~GeV,m_{\tilde{g}} = 250~GeV,
m_{\tilde{l}} = m_H = m_{\tilde{H}} = 400~GeV, m_{\tilde{q}} =
600~GeV, m_L = m_{\tilde{L}} = 900~GeV, m_Q = m_{\tilde{Q}} =
2.2~TeV$.  Using Eq. (9) and the input values of $G_F, m_Z$ and
$\alpha(0)$, this choice yields
$\tilde{\Delta}_i(\mu_0=20~TeV) = (1.26,1.40,1.24)$ and
$\hat{\alpha}_{1,2}(m_Z) = (1/59.56, 1/29.90)$ in the MDSP scheme 
\cite{confuse}.  Using these
and an input $\hat{\alpha}_3(m_Z) = 0.127$, we determine
$\alpha_i^{-1}(\mu_0)$ by means of 
Eqs. (6)-(8), which we use in turn to extrapolate to higher
values of $\mu$ with the help of Eqs. (2)-(4).  As can be seen
from Fig. 1, the three gauge couplings meet at a scale $M_X \approx
10^{17}~GeV$ (to within 2\% difference from each other), 
with a unified value of the gauge couplings $\alpha_X
\approx 0.24$.  

\underline{{\it Case 2:}}  $m_{\tilde{W}} = 75~GeV, m_{\tilde{g}} = 215~GeV,
m_{\tilde{l}} = m_H = m_{\tilde{H}} = 300~GeV, m_{\tilde{q}} = 500, m_L =
m_{\tilde{L}} = 500~GeV, m_Q=m_{\tilde{Q}}=1.5~TeV$:  For this case \cite{change},
the couplings  meet almost perfectly at $M_X \approx .8 \times 10^{17}~GeV$
with $\alpha_X \approx 0.25$ and $\hat{\alpha}_3(m_Z) = 0.125$ (see
Fig. 2).

\underline{{\it Case 3:}}  $m_{\tilde{W}} = 90~GeV, m_{\tilde{g}} =
170~GeV, m_{\tilde{l}} =
m_H = m_{\tilde{H}} = 400~GeV, m_{\tilde{q}} =600~GeV, m_L = m_{\tilde{L}} =
900~GeV, m_Q = m_{\tilde{Q}} = 2.2~TeV$.  Here we get 
perfect meeting with $M_X \approx .7 \times
10^{17}~GeV, \alpha_X \approx .22$ and $\hat{\alpha}_3(m_Z) = .123$
(see Fig. 3).

While we have not explored the parameter space pertaining to the
spectrum of the new particles and variation in
$\alpha_3(m_Z)$ in any detail, we find it indeed $remarkable$ that the
three couplings meet, even perfectly for many cases, for a fairly wide
variation in the ESSM spectrum beyond what we have exhibited here [30].  
The corresponding values of $\alpha_X, M_X$ and  $\hat{\alpha}_3(m_Z)$ in ESSM
are found to lie in the
ranges of [31]:  
\begin{equation}
\alpha_X \approx (.2-.3); M_X = (.7-1.2)\times
10^{17}~GeV, \hat{\alpha}_3(m_Z) = .122-.128~.  
\end{equation}
Thus we see that ESSM leads to coupling--unification, with an
intermediate value of $\alpha_X$, and a lower value of $\alpha_3(m_Z)$
than that needed for MSSM unification, just as desired.  The resulting
$M_X \sim 10^{17}~GeV$ is higher than the MSSM value, but it is
still lower than  the one--loop
string--unification scale of Ref. \cite{ginsparg}, which, for
$\alpha_X \approx 0.25$, yields $M_{\rm st} \approx 7 \times
10^{17}~GeV$.  This remaining gap between $M_X$ and
$M_{\rm st}$ may have its resolution in part due to the increased
importance of two--loop string threshold effects, corresponding to an
intermediate value of $\alpha_X$, which could lead to significant
corrections to the one--loop formula for $M_{\rm st}$
[6], and in part due to the relative imporance of three and higher
loop effects, which may shift $M_X$ (see remarks below).  In other
words, considering the proximity of $M_X \sim 10^{17}~GeV$ to the
expected string scale of $(5-8)\times 10^{17}~GeV$, contributions from
the infinite tower of heavy string-states, which have been neglected
in the running of $\alpha_i$'s, and quantum gravity may play an
important role in bridging the relatively small gap between $M_X$ and
$M_{\rm st}$ [32].  In summary, ESSM predicts (a) an intermediate value of
$\alpha_X$ which may help
stabilize the dilaton, (b)  
a value of $M_X \sim M_{\rm st} \ge 10^{17}~GeV$,
would fare better than the case of $M_{\rm st} \sim 2 \times
10^{16}~GeV$ [17] in avoiding the potential problem of rapid proton decay
induced through d = 5 operators and  (d) a lower
$\alpha_3(m_Z)$ than the case of MSSM.  
These appear to be distinct advantages of ESSM over MSSM.

Before concluding, the following points are worth noting.  \newline
\noindent (i) Even if
ESSM--unification might be closer to the truth, it provides a simple
reason why the couplings appear to meet, at least approximately, even
for MSSM.  As alluded to before, the reason is that in one loop,
unification of couplings in one scheme implies
that for the other, though with a vastly different $\alpha_X$.  The
two models differ only in two loop and thereby
in the resulting values of $M_X$, $\alpha_X$ as well as
$\alpha_3(m_Z)$.  \newline
\noindent (ii)  The
two loop gauge coupling contribution (i.e., $b_{ij}$ terms in Eq. (2))
which raise the slopes of $\alpha_i$, together with the softening
effects of the Yukawa contributions which do the opposite, turned out
to play an important role [33] in achieving unification for
ESSM.  It is the interplay of these two contributions
which leads to a good meeting of
the three gauge couplings (Fig. 1-3) with a low $\alpha_3(m_Z)$.  \newline
\noindent (iii) Although 3--loop effects could be
important especially in fixing $M_X$, we expect our calculation based
on 2--loop contributions presented here to be still fairly reliable,
at least for the range $m_Z \le \mu \le 10^{15}~GeV$ for which the 
couplings are small (i.e.,
$\alpha_{1,2} \le 0.12, \alpha_3 \le 0.18$, see Fig. 1-3).  By the
time $\mu$ rises to $10^{15}~GeV$, the three couplings,
especially $\alpha_1$ and $\alpha_2$, begin turning sharply upward
together in a manner that the tendency of the three curves to converge
to a common meeting point is already apparent (see Fig. 1-3).  Owing
to the coupled RGE for the three $\alpha_i$, we suspect
that this tendency would persist in three and higher loops [32].  \newline
\noindent (iv) A related remark: if $\alpha_X$
has an intermediate value, so that it may help stabilize the dilaton,
the relative importance of two and possibly higher loops near $M_X$, compared to
the case of MSSM, cannot be avoided.  Yet as shown here, unification
can already be seen quite visibly in two loops in the sense commented
above.  \newline
\noindent (v) This preliminary work of ESSM motivates further study of
the evolution of the gauge couplings, fermion and scalar masses as
well as of radiative symmetry breaking in ESSM with the inclusion of
two--loop evolution of Yukawa couplings and three--loop effects in
Eq. (2).  \newline
\noindent (vi) Last but not least, ESSM predicts [12-16]
two complete vector--like families with leptonic and quark
members having masses in the ranges of (200 GeV--1 TeV) and (500 GeV
--2.5 TeV) respectively.  Their mass--pattern, mixing and decay modes
as well as characteristic signals have been considered in detail in
Ref. [14].  To mention just a few such signals, pair production of
vector--like quarks at LHC and/or future version of SSC would lead to
systems such as $(b\overline{b} + 4Z+W^+W^-$) and $(b
\overline{b}+2Z+W^+W^-)$, while an $e^+e^-$ collider (NLC)
could produce $E^-E^+$ and even flavor--violating $N\overline{\nu}_\tau$
pair appreciably, followed by the decay $N \rightarrow Z+\nu_\tau
\rightarrow (e^+e^-)+\nu_\tau$.  Furthermore, once the relevant
momentum transfer for sub-processes exceeds about $m_Q$
in hadronic colliders, the corresponding $\alpha_3$ would grow
significantly due to contributions from virtual (or real) heavy quark
pairs and their SUSY partners.  This would manifest for example in
enhancement of JET cross sections, even below threshold for production
of real heavy quarks of a nature recently reported by the CDF group
[34].  Even though the CDF findings may or may not reflect truly new
physics, the phenomenon should reappear in high $p_T$--processes of
future colliders including the LHC if the vector--like quarks with
masses as above exist.

To conclude, ESSM, possessing two extra vector--like families 
with masses of order 1 TeV [35], provides (a) a simple explanation
of the inter--family mass--hierarchy [12,13] as well as (b) unification
with a higher $\alpha_X \sim .2-.3$, a higher $M_X \sim 10^{17}~GeV$ and
a lower $\alpha_3(m_Z)$ compared to MSSM [31], just as desired.  
The emergence of an extra pair of ${\bf 16}+
{\bf \overline{16}}$ is rather generic in string theories.  But the
derivation of the ESSM spectrum together with a standard model--like
gauge symmetry and Yukawa coupling matrix, as assumed here, from a
string theory remains an important task.  Owing to the advantages mentioned 
above, ESSM appears to be an attractive, yet falsifiable, alternative to MSSM.

\section*{Acknowledgments}

JCP acknowledges the kind hospitality which he received at
the Institute for Advanced Study and at CERN 
where part of this work was carried out.  
We would especially like to thank E. Witten for many
helpful discussions on string unification and for his
interest in this work.  We have greatly benefitted from discussions
with J. Bagger, L. Clavelli, S. Pokorski and especially M. Bastero-Gil on the issue
of low energy threshold effects.  We are also grateful to  R. Dick, K. Dienes, 
M. Dine, A. Faraggi, L. Ibanez, J.B. Kim, C. Kolda, P. Langacker, 
 J. March-Russell, A. Rasin, G. Ross, O.V. Tarasov, G. Veneziano 
and F. Wilczek for helpful comments.

\begin{figure}
\centering
\epsfysize=6in
\hspace*{0in}
\epsffile{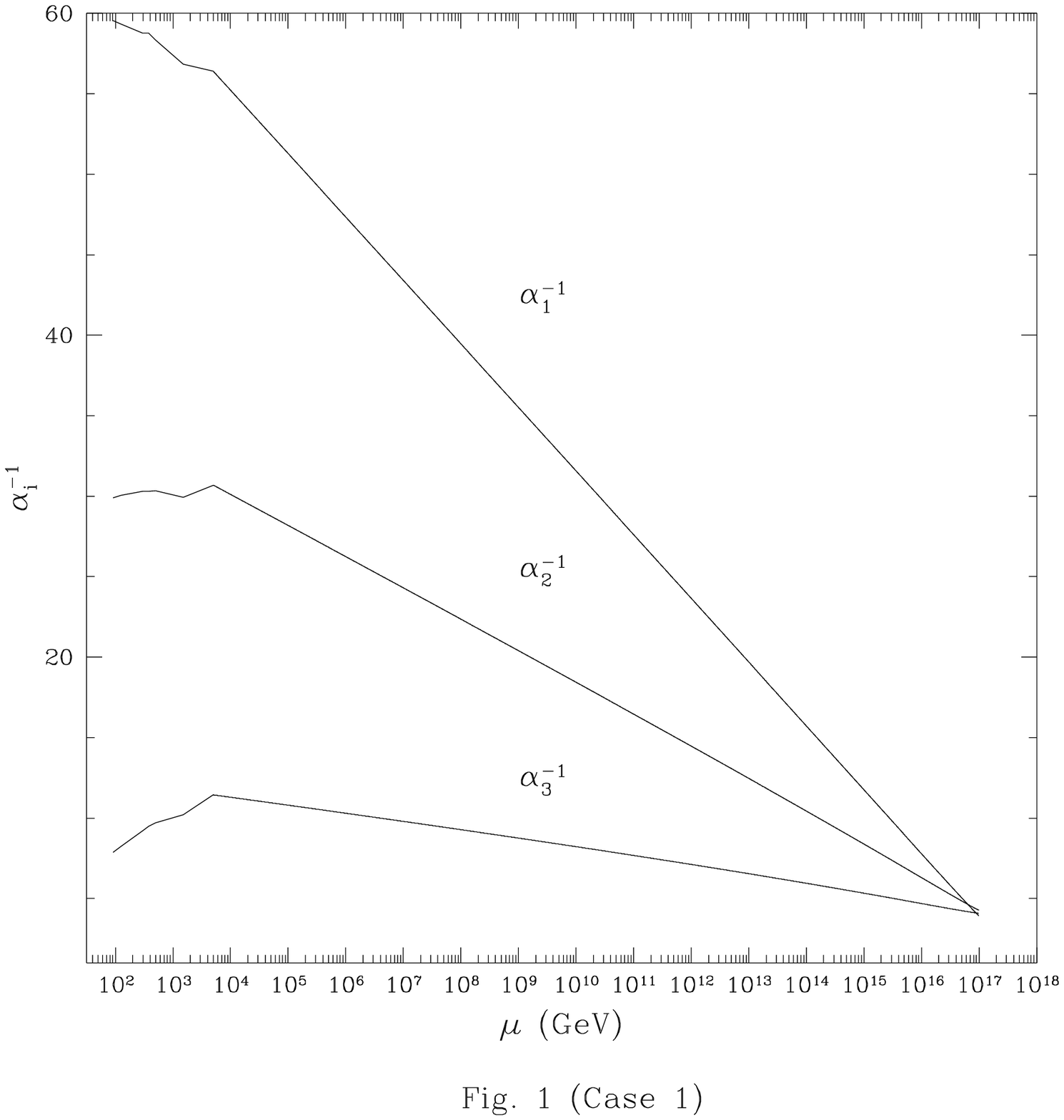}
\end{figure}
\newpage
\begin{figure}
\centering
\epsfysize=4in
\hspace*{0in}
\epsffile{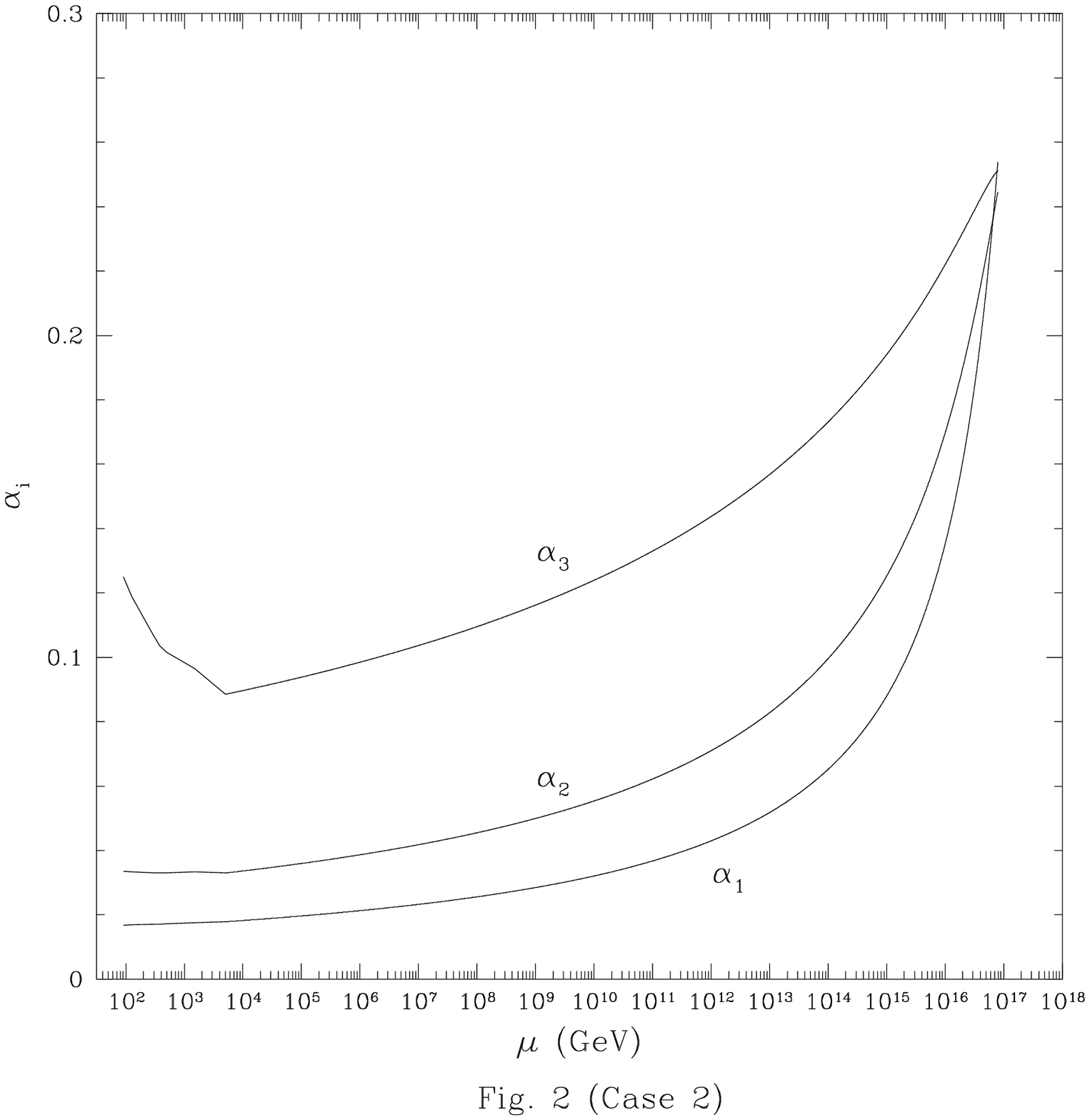}
\end{figure}
\begin{figure}
\centering
\epsfysize=4in
\hspace*{0in}
\epsffile{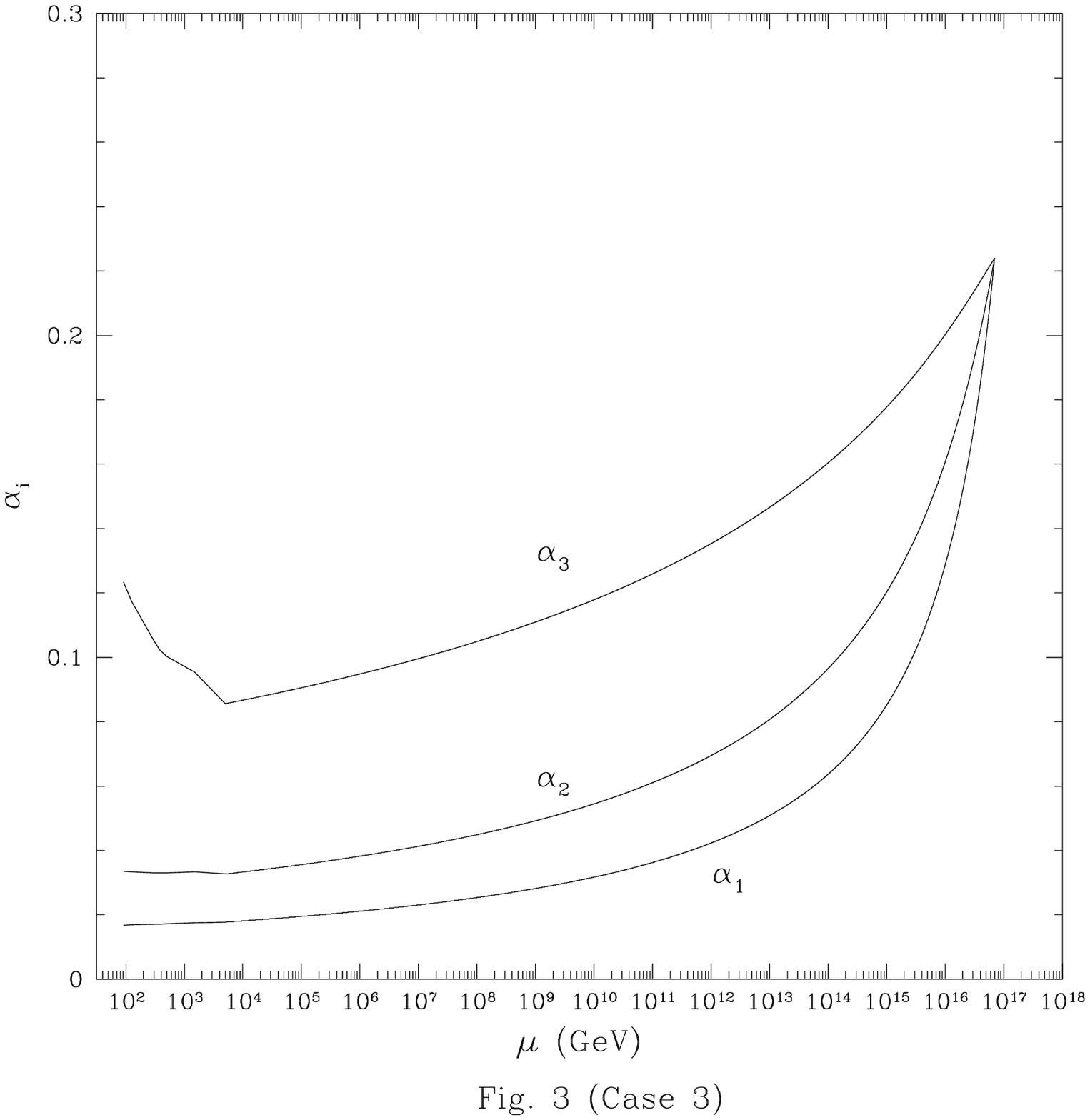}
\end{figure}


\begin{thebibliography}{99}
\bibitem{25}
         It is instructive to note that for $m_Z \le m \le \mu$,
$\Delta_i^F = (\hat{b}_i/2\pi)[-{\rm
ln}(\mu/m)+5/6+1/10(m_Z/m)^2-3(m^2/\mu^2)+O\{(m^4/\mu^4){\rm ln}(\mu/m),(m_Z^4/m^4)\},
]$.  This shows that for
$m \sim m_Z$ or $\mu$, the leading log--term gives a
poor approximation.
\bibitem{virtue}
        This also has the virtue that our results will be essentially
free from the arbitrariness in the
detailed choice of the Yukawa couplings.
\bibitem{change}

       The small variations ($\le 0.3\%$) in $\hat{\alpha}_{1,2}(m_Z)$
due to change in spectra in going from Case 1 to Case 2 
will be ignored in this note.
\bibitem{imp}
   
       Since string-threshold effects are expected to introduce 
corrections to the three $\alpha_i$'s, differing by a few percent
near $M_X$ [19], we regard cases of almost
perfect meeting (Fig. 2-3) as being on par with those for which the
meeting occurs to within (2-3)\% (as in Fig. 1).
\bibitem{pqr}
       
       Including the $-8\%$ difference between
$\alpha_3(m_Z)|_{\overline{MS}}$ and $\hat{\alpha}_3(m_Z)$
\cite{confuse}, the values of $\hat{\alpha}_3(m_Z) \approx .122-.128$
in the MDSP scheme needed for unification in ESSM would correspond to
$\alpha_3(m_Z)|_{\overline{MS}} \approx .112-.118$.  This should be 
compared with
the value $ \alpha_3(m_Z)_{\overline{MS}} \ge .127$ needed for 
MSSM unification, barring GUT and string threshold
effects, on the one hand, and the world average
value of $.117 \pm .005$ \cite{pdg}, on the other hand.  Both ESSM and
MSSM values are still subject to string and possibly
unification--scale threshold effects of a few percent [19].
\bibitem{32}
We are tempted to conjecture that with $\alpha_X \approx .2-.3$, the
increased importance of the contribution from the heavy string tower
of states might in fact lead to a sharp slowing down in the growth of
the $\alpha_i$'s as $\mu$ exceeds $10^{17}~GeV$ (compare with Figs. 
1-3), so that they may approach their unified fixed point value (with
$\beta_i \rightarrow 0$) as $\mu$ reaches or exceeds $M_{\rm st}$.
Such a behavior, although hard to demonstrate with the present state
of the art, would seem to be in accord with the intrinsic finiteness
of the string theory.  The true $M_X$ and $\alpha_{\rm st}$ 
in this case would be somewhat higher
than that exhibited by the meeting point in Fig. 1-3.
\bibitem{abc}

The importance of two--loop contributions in ESSM arises espcially
because the one--loop $\beta$--function for $\alpha_3$ almost
vanishes, i.e, $b_3=-9+2n_g=+1$ for ESSM, but $-3$ for MSSM, see
Eq. (2).
\bibitem{inter}
      Although in this paper we have assumed the standard model gauge
symmetry below the string scale, the ESSM spectrum would still help
raise $\alpha_X$ even if a higher symmetry like $G_{224}\equiv SU(2)_L \times
SU(2)_R \times SU(4)_C$ or one of its subgroups 
emerges from strings, which breaks to the
standard model symmetry at a scale $\sim 10^{16}~GeV$.  These cases
need to be studied separately.  

\section*{Figure Caption}

Figs. 1-3: Plots of $\alpha_i^{-1}$ (Fig 1) and $\alpha_i$ (Figs. 2-3) as a 
function of $\mu$.
\end{thebibliography}
\end{document}